\documentstyle[preprint,aps]{revtex}

\begin{document}
\title{{\bf Bound states of a more general exponential screened Coulomb potential }}
\author{Sameer M. Ikhdair\thanks{%
sikhdair@neu.edu.tr} and \ Ramazan Sever\thanks{%
sever@metu.edu.tr}}
\address{$^{\ast }$Department of Physics, \ Near East University, Nicosia, North
Cyprus, Mersin 10, Turkey\\
$^{\dagger }$Department of Physics, Middle East Technical University, 06531
Ankara, Turkey.}
\date{\today
}
\maketitle

\begin{abstract}
An alternative approximation scheme has been used in solving the
Schr\"{o}dinger equation to the more general case of exponential screened
Coulomb potential, $V(r)=-(a/r)\left[ 1+(1+br)e^{-2br}\right] .$ The bound
state energ\i es of the $1s,$ $2s,$ and $3s-$states, together with the
ground state wave function are obtained analytically upto the second
perturbation term.


PACS\ NO: 03.65.Ge
\end{abstract}


\section{Introduction}

\noindent A more general exponential screened Coulomb (MGESC) potential of
the form:
\begin{equation}
V\left( r\right) =-\left( \frac{a}{r}\right) \left[ 1+(1+br)\exp (-2br)%
\right] ,
\end{equation}
where $a$ is the strength coupling constant and $b$ is the screening
parameter, is known to describe adequately the effective interaction in
many-body enviroment of a variety of fields such as atomic, nuclear,
solid-state, plasma physics and quantum field theory [1,2]. It is also used
in describing the potential between an ionized impurity and an electron in a
metal [3,4] or a semiconductor \ [5] and the electron-positron interaction
in a positronium atom in a solid [6].

The Schr\"{o}dinger equation for such a potential does not admit
exact solutions, various approximate methods both numerical [7] and
analytical [8] have been developed \ The MGESC Potential defined for
an electron of the helium atom in the field of other electrons and
nucleus has been investigated by Gerry and Laub [9]. Further, the
large-N expansion was used to obtain the bound-state energy of the
ground state and the first excited state and the corresponding wave
functions analytically by Ref.[10].

In this paper, we calculate the binding energy eigenvalues of MGESC
potential using a novel perturbative formalism [11] which has recently been
used in solving the Schr\"{o}dinger equation to obtain the bound-state
energies as well as the wave functions for different types of potentials
[11,12] in both bound and continuum regions. This novel treatment is based
on the decomposition of the radial Schr\"{o}dinger equation into two pieces
having an exactly solvable part with an addi\i tional piece leading to
either a closed analytical solution or approximate treatment depending on
the nature of the perturbed potential.

The contents of this paper is as follows. In Section \ref{TM} we breifly
outline the method with all necessary formulae to perform the current
calculations. In Section \ref{A} we apply this method to the Schr\"{o}dinger
equation using a more general exponential screening Coulomb potential to
obtain analytical expressions for the bound-state energy and eigen functions
of different energy states. Finally, we end with some results and
conclusions in Section \ref{RAC}.

\section{The Method}

\label{TM}For a spherically symmetric potential, the corresponding
Schr\"{o}dinger equation, in the bound state domain, with the radial wave
function reads

\begin{equation}
\frac{\hbar ^{2}}{2m}\frac{\psi _{n}^{\prime \prime }(r)}{\psi _{n}\left(
r\right) }=V(r)-E_{n},
\end{equation}
with

\begin{equation}
V\left( r\right) =\left[ V_{0}(r)+\frac{\hbar ^{2}}{2m}\frac{\ell (\ell +1)}{%
r^{2}}\right] +\Delta V(r),
\end{equation}
where $\Delta V(r)$ is a perturbing potential and $\psi _{n}(r)=\chi
_{n}(r)u_{n}(r)$ is the full radial wave function, in which $\chi _{n}(r)$
is the known normalized wave function of the unperturbed Schr\"{o}dinger
equation whereas $u_{n}(r)$ is a moderating wave function corresponding to
the perturbing potential. Following the method given in Refs. [11,12], we
may rewrite (2) as

\begin{equation}
\frac{\hbar ^{2}}{2m}\left( \frac{\chi _{n}^{\prime \prime }(r)}{\chi _{n}(r)%
}+\frac{u_{n}^{\prime \prime }(r)}{u_{n}(r)}+2\frac{\chi _{n}^{\prime
}(r)u_{n}^{\prime }(r)}{\chi _{n}(r)u_{n}(r)}\right) =V(r)-E_{n}.
\end{equation}
The logarithmic derivatives of the unperturbed $\chi _{n}(r)$ and perturbed $%
u_{n}(r)$ wave functions are given by

\begin{equation}
W_{n}(r)=-\frac{\hbar }{\sqrt{2m}}\frac{\chi _{n}^{\prime }(r)}{\chi _{n}(r)}%
\text{ \ \ and \ \ }\Delta W_{n}=-\frac{\hbar }{\sqrt{2m}}\frac{%
u_{n}^{\prime }(r)}{u_{n}(r)},
\end{equation}
which leads to

\begin{equation}
\frac{\hbar ^{2}}{2m}\frac{\chi _{n}^{\prime \prime }(r)}{\chi _{n}(r)}%
=W_{n}^{2}(r)-\frac{\hbar }{\sqrt{2m}}W_{n}^{^{\prime }}(r)=\left[ V_{0}(r)+%
\frac{\hbar ^{2}}{2m}\frac{\ell (\ell +1)}{r^{2}}\right] -\varepsilon _{n},
\end{equation}
where $\varepsilon _{n}$ is the eigenvalue for the exactly solvable
potential of interest, and

\begin{equation}
\frac{\hbar ^{2}}{2m}\left( \frac{u_{n}^{\prime \prime }(r)}{u_{n}(r)}+2%
\frac{\chi _{n}^{\prime }(r)u_{n}^{\prime }(r)}{\chi _{n}(r)u_{n}(r)}\right)
=\Delta W_{n}^{2}(r)-\frac{\hbar }{\sqrt{2m}}\Delta W_{n}^{\prime
}(r)+2W_{n}(r)\Delta W_{n}(r)=\Delta V(r)-\Delta \varepsilon _{n},
\end{equation}
in which $\Delta \varepsilon _{n}=E_{n}^{(1)}+E_{n}^{(2)}+\cdots $ is the
correction term to the energy due to $\Delta V(r)$ and $E_{n}=\varepsilon
_{n}+\Delta \varepsilon _{n}.$ If Eq. (7), which is the most significant
piece of the present formalism, can be solved analytically as in (6), then
the whole problem, in Eq. (2) reduces to the following form

\begin{equation}
\left[ W_{n}(r)+\Delta W_{n}(r)\right] ^{2}-\frac{\hbar }{\sqrt{2m}}%
(W_{n}(r)+\Delta W_{n}(r))^{\prime }=V(r)-E_{n},
\end{equation}
which is a well known treatment within the frame of supersymmetric quantum
theory (SSQT) [13]. Thus, if the whole spectrum and corresponding
eigenfunctions of the unperturbed interaction potential are known, then one
can easily calculate the required superpotential $W_{n}(r)$ for any state of
interest leading to direct computation of related corrections to the
unperturbed energy and wave function.

For the perturbation technique, we can split the given potential in Eq.(2)
into two parts. The main part corresponds to a shape invariant potential,
Eq. (6), for which the superpotential is known analytically and the
remaining part is treated as a perturbation, Eq. (7). Therefore, the MGESC
potential can be solved using this method. In this case, the zeroth-order
term corresponds to the Coulomb potential while higher-order terms consitute
the perturbation. However, the perturbation term in its present form cannot
be solved exactly through Eq. (7). Thus, one should expand the functions
related to the perturbation in terms of the perturbation parameter $b$,

\begin{equation}
\Delta V(r;b)=\sum_{i=1}^{\infty }b_{i}V_{i}(r),\text{ \ \ }\Delta
W_{n}(r;b)=\sum_{i=1}^{\infty }b_{i}W_{n}^{(i)}(r),\text{ \ }%
E_{n}^{(i)}(b)=\sum_{i=1}^{\infty }b_{i}E_{n}^{(i)},
\end{equation}
where $i$ denotes the perturbation order. Substitution of the above
expansions into Eq. (7) and equating terms with the same power of $b$\ on
both sides up to $O(b^{3})$ yields

\begin{equation}
2W_{n}(r)W_{n}^{(1)}(r)-\frac{\hbar }{\sqrt{2m}}\frac{dW_{n}^{(1)}(r)}{dr}%
=V_{1}(r)-E_{n}^{(1)},
\end{equation}

\begin{equation}
W_{n}^{(1)2}(r)+2W_{n}(r)W_{n}^{(2)}(r)-\frac{\hbar }{\sqrt{2m}}\frac{%
dW_{n}^{(2)}(r)}{dr}=V_{2}(r)-E_{n}^{(2)},
\end{equation}

\begin{equation}
2\left[ W_{n}(r)W_{n}^{(3)}(r)+W_{n}^{(1)}(r)W_{n}^{(2)}(r)\right] -\frac{%
\hbar }{\sqrt{2m}}\frac{dW_{n}^{(3)}(r)}{dr}=V_{3}(r)-E_{n}^{(3)}.
\end{equation}
Hence, unlike the other perturbation theories, Eq. (7) and its expansion,
Eqs. (10)-(12), give a flexibility for the easy calculations of the
perturbative corrections to energy and wave functions for the $nth$ state of
interest through an appropriately chosen perturbed superpotential.

\section{Application to the MGESC Potential}

\label{A}Considering the recent interest in various power-law potentials in
the literature, we work throughout the article within the frame of low
screening parameter $b$. In this regard, the MGESC potential can be expanded
in power series of the screening parameter $b$ as [10,12]

\begin{equation}
V(r)=-\left( \frac{a}{r}\right) \left[ 1+\left( 1+br\right) \exp (-2br)%
\right] =-\frac{a}{r}-\frac{a}{r}\sum_{i=0}^{\infty }V_{i}(br)^{i},
\end{equation}
where the perturbation coefficients $V_{i}$ are given by

\begin{equation}
V_{1}=-1,\text{ }V_{2}=0,\text{ }V_{3}=2/3,\text{ }V_{4}=-4/6,\text{ }%
V_{5}=12/30,\cdots .
\end{equation}
Therefore, we apply this approximation method to the MGESC potential with
the angular momentum barrier

\begin{equation}
V(r)=-\left( \frac{a}{r}\right) \left[ 1+\left( 1+br\right) e^{-2br}\right] +%
\frac{\ell (\ell +1)\hbar ^{2}}{2mr^{2}}=\left[ V_{0}(r)+\frac{\ell (\ell
+1)\hbar ^{2}}{2mr^{2}}\right] +\Delta V(r),
\end{equation}
where the first piece is the shape invariant zeroth-order which is an
exactly solvable piece corresponding to the unperturbed Coulomb potential
with $V_{0}(r)=-2a/r$ while $\Delta
V(r)=ab-(2ab^{3}/3)r^{2}+(4ab^{4}/6)r^{3}-(12ab^{5}/30)r^{4}+\cdots $ is the
perturbation term. The literature is rich with examples of particular
solutions for such power-law potentials employed in different fields of
physics, for recent applications see Refs. [14,15]. At this stage one may
wonder why the series expansion is truncated at a lower order. This can be
understood as follows. It is widely appreciated that convergence is not an
important or even desirable property for series approximations in physical
problems. Specifically, a slowly convergent approximation which requires
many terms to achieve reasonable accuracy is much less valuable than the
divergent series which gives accurate answers in a few terms. This is
clearly the case for the MGESC problem [16]. However, it is worthwhile to
note that the main contributions come from the first three terms. Thereby,
the present calculations are performed upto the second-order involving only
these additional potential terms, which suprisingly provide highly accurate
results for small screening parameter $b.$

\subsection{Ground State Calculations $\left( n=0\right) $}

In the light of Eq. (6), the zeroth-order calculations leading to exact
solutions can be carried out readily by setting the ground-state
superpotential and the unperturbed exact energy as

\begin{equation}
W_{n=0}\left( r\right) =-\frac{\hbar }{\sqrt{2m}}\ \frac{\ell +1}{r}+\sqrt{2m%
}\frac{a}{(\ell +1)\hbar },\text{ \ \ }E_{n}^{(0)}=-\frac{2ma^{2}}{\hbar
^{2}(n+\ell +1)^{2}},\text{ \ \ \ }n=0,1,2,....
\end{equation}
and from the literature, the corresponding normalized Coulomb bound-state
wave function [17]

\begin{equation}
\chi _{n}(r)=N_{n,l}^{(C)}r^{\ell +1}\exp \left[ -\beta r\right] \times
L_{n}^{2\ell +1}\left[ 2\beta r\right] ,
\end{equation}
in which $N_{n,l}^{(C)}=\left[ \frac{4ma}{\left( n+\ell +1\right) \hbar ^{2}}%
\right] ^{\ell +1}\frac{1}{(n+\ell +1)}\frac{1}{\sqrt{\frac{\hbar ^{2}}{2man!%
}(n+2\ell +1)!}}$ is a normalized constant,\ \ $\beta =\frac{2ma}{\left(
n+\ell +1\right) \hbar ^{2}}$ and $L_{n}^{k}\left( x\right)
=\sum_{m=0}^{n}(-1)^{m}\frac{(n+k)!}{\left( n-m\right) !(m+k)!m!}x^{m}$ is
an associate Laguarre polynomial function [18].

For the calculation of corrections to the zeroth-order energy and wave
function, one needs to consider the expressions leading to the first- and
second-order perturbation given by Eqs. (10)--(12). Multiplication of each
term in these equations by $\chi _{n}^{2}(r)$, and bearing in mind the
superpotentials given in Eq. (5), one can obtain the straightforward
expressions for the first-order correction to the energy and its
superpotential:
\begin{equation}
E_{n}^{(1)}=\int_{-\infty }^{\infty }\chi _{n}^{2}(r)\left( -\frac{2ab^{3}}{3%
}r^{2}\right) dr,\text{ }W_{n}^{(1)}\left( r\right) =\frac{\sqrt{2m}}{\hbar }%
\frac{1}{^{X_{n}^{2}(r)}}\int^{r}\chi _{n}^{2}(x)\left[ E_{n}^{(1)}+\frac{%
2ab^{3}}{3}x^{2}\right] dx,\
\end{equation}
and also for the second-order correction and its superpotential:

\[
E_{n}^{(2)}=\int_{-\infty }^{\infty }\chi _{n}^{2}(r)\left[ \frac{4ab^{4}}{6}%
r^{3}-W_{n}^{(1)2}\left( r\right) \right] dr,\text{ }
\]
\begin{equation}
W_{n}^{(2)}\left( r\right) =\frac{\sqrt{2m}}{\hbar }\frac{1}{^{X_{n}^{2}(r)}}%
\int^{r}\chi _{n}^{2}(x)\left[ E_{n}^{(2)}+W_{n}^{(1)2}(x)-\frac{4ab^{4}}{6}%
x^{3}\right] dx\ ,
\end{equation}
for any state of interest. The above expressions calculate $W_{n}^{(1)}(r)$
and $W_{n}^{(2)}(r)$\ explicitly from the energy corrections $E_{n}^{(1)}$
and $E_{n}^{(2)}$ respectively, which are in turn used to calculate the
moderating wave function $u_{n}(r).$

Thus, through the use of Eqs. (18) and (19), after some lengthy and tedious
integrals, we find the zeeroth order energy shift and their moderating
superpotentials as

\[
E_{0}^{(1)}\ =-\frac{\hbar ^{4}\left( \ell +1\right) ^{2}\left( \ell
+2\right) \left( 2\ell +3\right) }{12am^{2}}b^{3},
\]

\begin{eqnarray*}
E_{0}^{(2)} &=&\frac{\hbar ^{6}\left( \ell +1\right) ^{3}\left( \ell
+2\right) \left( 2\ell +3\right) \left( 2\ell +5\right) }{48a^{2}m^{3}}b^{4}
\\
&&-\frac{\hbar ^{10}\left( \ell +1\right) ^{6}\left( \ell +2\right) \left(
2\ell +3\right) \left( 8\ell ^{2}+37\ell +43\right) }{1152a^{4}m^{5}}b^{6},
\end{eqnarray*}

\[
W_{0}^{(1)}(r)=-\frac{\hbar \left( \ell +1\right) b^{3}r}{3\sqrt{2m}}\left[
r-\frac{\hbar ^{2}\left( \ell +1\right) \left( \ell +2\right) }{2am}\right]
,
\]
\begin{equation}
W_{0}^{(2)}(r)=-\frac{\hbar b^{4}a_{3}r}{2\sqrt{2m}}\left\{
b^{2}r^{3}+a_{1}r^{2}+a_{2}\left[ r+\frac{\hbar ^{2}(\ell +1)(\ell +2)}{2am}%
\right] \right\} -\frac{\hbar \left( \ell +1\right) }{2\sqrt{2m}a}%
E_{0}^{(2)},
\end{equation}
in which

\begin{eqnarray}
a_{1} &=&\frac{\hbar ^{2}(\ell +1)(3\ell +7)b^{2}}{2am}-\frac{12am}{\hbar
^{2}(\ell +1)^{2}},\text{ \ }a_{2}=\left[ \frac{\hbar ^{4}(\ell
+1)^{2}(8\ell ^{2}+37\ell +43)b^{2}}{8a^{2}m^{2}}-\frac{3}{2}\frac{(2\ell +5)%
}{(\ell +1)}\right] ,\text{ }  \nonumber \\
a_{3} &=&\frac{\hbar ^{2}(\ell +1)^{3}}{18am}
\end{eqnarray}
Therefore, setting $\beta =b/a,$ the analytical expression for the ground $s$%
-state energy is explicitly given, to order $\beta ^{6},$ in atomic units $%
(\hbar =m=1):$

\begin{equation}
E_{0}/a^{2}=-2+\beta -\frac{1}{2}\beta ^{3}+\frac{5}{8}\beta ^{4}-\frac{43}{%
192}\beta ^{6}+\cdots ,\text{ }
\end{equation}
and the full radial wavefunction is given by

\begin{equation}
\psi _{n=0,\ell }(r)\approx \chi _{n=0,\ell }\exp \left( -\frac{\sqrt{2m}}{%
\hbar }\int^{r}\left( W_{0}^{(1)}\left( x\right) +W_{0}^{(2)}\left( x\right)
\right) dx\right) .
\end{equation}
Hence, the explicit form of the full wave function in (23) for the ground
state is

\begin{equation}
\psi _{n=0,\ell }(r)=\left[ \frac{4ma}{(\ell +1)\hbar ^{2}}\right] ^{\ell +1}%
\frac{1}{(\ell +1)^{2}}\sqrt{\frac{2am}{\hbar ^{2}(2\ell +1)!}}r^{\ell
+1}\exp (P(r)),
\end{equation}
with $P(r)=\sum_{i=1}^{5}p_{i}r^{i}$ is a polynomial of fifth order having
the following coefficients:
\[
p_{1}=\frac{(\ell +1)}{2a}E_{0}^{(2)}-\frac{2am}{(\ell +1)\hbar ^{2}},\text{
\ }p_{2}=\frac{9}{4}\frac{(\ell +2)}{(\ell +1)^{2}}a_{3}^{2}a_{4}b^{4},\text{
}
\]
\begin{equation}
\text{\ }p_{3}=\frac{1}{6}a_{3}a_{4}b^{4},\text{ }p_{4}=\frac{1}{8}%
a_{1}a_{3}b^{4},\text{ }p_{5}=\frac{1}{10}a_{3}b^{6},\text{\ }
\end{equation}
where $a_{4}=a_{2}+\frac{12am}{\hbar ^{2}(\ell +1)^{2}b}.$

\subsection{Excited state calculations $(n\geq 1)$}

The calculations lead to a handy recursion relations in the case of ground
states, however it becomes extremely cumbersome in the description of radial
excitations when nodes of wavefunctions are taken into account, in
particular during the higher order calculations. Although several attempts
have been made to bypass this difficulty and improve calculations in dealing
with excited states, (cf. e.g. [19], and the references therein) within the
frame of SSQM.

Using Eqs. (5) and (17), the superpotential $W_{n}(r)$ which is related to
the excited states can be readily calculated by means of Eqs. (18) and (19).
Hence, the first-order corrections in the first excited state $(n=1)$ are

\[
E_{1}^{(1)}=-\frac{\hbar ^{4}\left( \ell +2\right) ^{2}\left( \ell +7\right)
\left( 2\ell +3\right) }{12am^{2}}b^{3},
\]
\

\begin{equation}
W_{1}^{(1)}(r)=-\frac{\hbar \left( \ell +2\right) b^{3}r}{3\sqrt{2m}}\left[
r+\frac{\hbar ^{2}(\ell +2)(\ell +3)}{2am}\right] .
\end{equation}
Consequently, substitution of the last equation into Eq. (19) allows us to
write down\

\begin{eqnarray}
\ E_{1}^{(2)} &=&\frac{\hbar ^{6}\left( \ell +2\right) ^{3}\left( \ell
+11\right) \left( 2\ell +3\right) \left( 2\ell +5\right) }{48a^{2}m^{3}}b^{4}
\nonumber \\
&&-\frac{\hbar ^{10}\left( \ell +2\right) ^{6}\left( \ell +3\right) \left(
2\ell +3\right) \left( 7\ell ^{2}+101\ell +211\right) }{1152a^{4}m^{5}}b^{6}.
\end{eqnarray}
Therefore, the analytical expressions for the first-excited $s$-state energy
is explicitly given, to order $\beta ^{6},$ in atomic units:

\begin{equation}
E_{1}/a^{2}=-\frac{1}{2}+\beta -7\beta ^{3}+\frac{55}{2}\beta ^{4}-\frac{211%
}{2}\beta ^{6}+\cdots .\text{ }
\end{equation}

The related radial wavefunction can be expressed in an analytical form by
means of Eqs (18), (19) and (23), if required. The appromation used in this
work would not affect considerably the sensitivity of the calculations. On
the other hand, it is found analytically that our investigations put forward
an interesting hierarchy between $W_{n}^{(1)}(r)$ terms of different quantum
states in the first order after circumventing the nodal difficulties
elegantly,\ \ \

\begin{equation}
W_{n}^{(1)}(r)=-\frac{\hbar \left( n+\ell +1\right) b^{3}r}{3\sqrt{2m}}\left[
r+\frac{\hbar ^{2}(n+\ell +1)(n+\ell +2)}{2am}\right] ,
\end{equation}
which, for the second excited state $\left( n=2\right) $ leads to the
first-order correction

\[
\ E_{2}^{(1)}=-\frac{\hbar ^{4}\left( \ell +3\right) ^{2}\left( \ell
+2\right) \left( 2\ell +23\right) }{12am^{2}}b^{3},
\]

\begin{equation}
W_{2}^{(1)}(r)=-\frac{\hbar \left( \ell +3\right) b^{3}r}{3\sqrt{2m}}\left[
r+\frac{\hbar ^{2}(\ell +3)(\ell +4)}{2am}\right] .
\end{equation}
Hence, substituting $W_{2}^{(1)}(r)$ into Eq.(19) gives the energy
correction in the second-order as\

\begin{eqnarray}
\ E_{2}^{(2)} &=&\frac{\hbar ^{6}\left( \ell +2\right) \left( \ell +3\right)
^{2}\left( 2\ell +5\right) \left( 2\ell ^{2}+45\ell +153\right) }{%
48a^{2}m^{3}}b^{4}  \nonumber \\
&&-\frac{\hbar ^{10}\left( \ell +2\right) \left( \ell +3\right) ^{5}(16\ell
^{4}+474\ell ^{3}+3879\ell ^{2}+12118\ell +12873)}{1152a^{4}m^{5}}b^{6}.
\end{eqnarray}
Therefore, the analytical expressions for the second-excited $s-$state
energy, to order $\beta ^{6},$ in atomic units:

\begin{equation}
E_{2}/a^{2}=-\frac{2}{9}+\beta -\frac{69}{2}\beta ^{3}+\frac{2295}{8}\beta
^{4}-\frac{347571}{64}\beta ^{6}+\cdots .\text{ }
\end{equation}

\section{Results And Conclusions}

\label{RAC}Some numerical values of the perturbed energies of the $1s,$ $2s$
and $3s$ states, in the atomic units, for various values of $\beta $ in the
range $0\leq \beta \leq 1.0$ are presented in Table 1. Our results are
consistent to order $\beta ^{6\text{ }}$with earlier results obtained by
applying the large $N$-expansion method [10]. Therefore, our results are
found in high agreement for small values of $\beta $ with those given in
[10]. On the other hand, for large screening parameter values, the accuracy
of our results exceeds the ones given before in Ref.[10]. These results also
appear to be in close agreement with the results obtained by solving Schr%
\"{o}dinger equation numerically with the same potential via Numerov's
method [10]. However, these results tend to deviate slightly as $\beta $
approaches $1.0$ with the results obtained from the numerical solution of
Schr\"{o}dinger equation. Moreover, we illustrate the improvement of energy
with respect to orders of $\beta $ in Table 2.

In this work, we have shown that the bound-state energies of a MGESC
potential for all eigenstates can be accurately determined within the
framework of a novel treatment. Avoiding the disadvantages of the standard
non-relativistic perturbation theories, the obtained expressions in the
present work have much simple forms than the ones shown in the previous work
[10].

Finally, the application of the present technique to MGESC potential for the
first time is really of great interest providing leading to analytical
expressions for both energy eigenvalues and wave functions which are likely
of much interest in different field of physics.

\acknowledgments S.M. Ikhdair wishes to dedicate this work to his
family for their love and assistance. This research was partially
supported by the Scientific and Technological Research Council of
Turkey.

\bigskip

\begin{table}[tbp]
\caption{Calculated binding energy eigenvalues for $0<\protect\beta \leq 1.0$
up to order $\protect\beta ^{6}.$}
\begin{tabular}{llllllll}
$\beta $ & $E_{00}/a^{2}$\tablenotemark[1]\tablenotetext[1]{Calculations to
order $\beta^{6}$.} & $E_{00}/a^{2}$ [10]\tablenotemark[2]%
\tablenotetext[2]{Calculations to order $\beta^{4}$.} & Numerical [10] & $%
E_{10}/a^{2}$\tablenotemark[1] & $E_{10}/a^{2}$ [10]\tablenotemark[2] &
Numerical [10] & $E_{20}/a^{2}$\tablenotemark[1] \\
\tableline$0.01$ & $-1.9900005$ &  &  & $-0.4900067$ &  &  & $-0.2122538$ \\
$0.02$ & $-1.9800039$ & $-1.98000$ & $(-1.98000)$ & $-0.4800516$ & $-0.48005$
& $(-0.48000)$ & $-0.2024526$ \\
$0.03$ & $-1.9700130$ &  &  & $-0.4701668$ &  &  &  \\
$0.04$ & $-1.9600304$ & $-1.96003$ & $(-1.96003)$ & $-0.460378$ & $-0.46038$
& $(0.46033)$ & $-0.183718$ \\
$0.05$ & $-1.9500586$ & $-1.95006$ &  & $-0.4507047$ & $-0.45070$ &  &  \\
$0.06$ & $-1.9401161$ & $-1.94010$ & $(-1.94010)$ & $-0.4411605$ & $-0.44116$
& $(0.44115)$ & $-0.1662097$ \\
$0.07$ & $-1.9301565$ &  &  & $-0.4317531$ &  &  &  \\
$0.08$ & $-1.9202305$ & $-1.92023$ & $(-1.92023)$ & $-0.4224852$ & $-0.42246$
& $(-0.4221)$ & $-0.1495594$ \\
$0.09$ & $-1.9103236$ &  &  & $-0.4133547$ &  &  &  \\
$0.10$ & $-1.9004377$ & $-1.90044$ & $(-1.90044)$ & $-0.4043555$ & $-0.4043$
& $(-0.4048)$ & $-0.1334655$ \\
$0.20$ & $-1.8030143$ & $-1.803$ &  & $-0.318752$ & $-0.312$ &  &  \\
$0.30$ & $-1.7086008$ & $-1.70844$ & $(-1.70958)$ & $-0.2431595$ & $.$ & $%
(-0.274)$ &  \\
$0.40$ & $-1.6169173$ &  &  &  &  &  &  \\
$0.50$ & $-1.5269368$ & $-1.523$ & $(-1.537)$ &  &  &  &  \\
$0.60$ & $-1.4374490$ &  &  &  &  &  &  \\
$0.70$ & $-1.3477860$ & $-1.321$ & $(-1.384)$ &  &  &  &  \\
$0.80$ & $-1.2587093$ & $-1.2$ &  &  &  &  &  \\
$0.90$ & $-1.1734581$ &  &  &  &  &  &  \\
$1.00$ & $-1.0989583$ & $-0.875$ & $(-1.194)$ &  &  &  &
\end{tabular}
\end{table}

\begin{table}[tbp]
\caption{Improvement in energy with respect to orders of $b/a.$}
\begin{tabular}{llllll}
$\beta $ & $(E_{00}/a^{2})_{0}$ & $(E_{00}/a^{2})_{1}$ & $(E_{00}/a^{2})_{3}$
& $(E_{00}/a^{2})_{4}$ & $(E_{00}/a^{2})_{6}$ \\
\tableline$0.02$ & $-2.0$ & $-1.98$ & $-1.980004$ & $-1.9800039$ & $%
-1.9800039$ \\
$0.05$ & $-2.0$ & $-1.95$ & $-1.9500625$ & $-1.9500586$ & $-1.9500586$ \\
$0.08$ & $-2.0$ & $-1.92$ & $-1.920256$ & $-1.9202304$ & $-1.9202305$ \\
$0.20$ & $-2.0$ & $-1.80$ & $-1.804$ & $-1.803$ & $-1.8030143$ \\
$0.50$ & $-2.0$ & $-1.50$ & $-1.5625$ & $-1.5234375$ & $-1.5269368$ \\
$0.80$ & $-2.0$ & $-1.20$ & $-1.456$ & $-1.20$ & $-1.2587093$ \\
$\beta $ & $(E_{10}/a^{2})_{0}$ & $(E_{10}/a^{2})_{1}$ & $(E_{10}/a^{2})_{3}$
& $(E_{10}/a^{2})_{4}$ & $(E_{10}/a^{2})_{6}$ \\
\tableline$0.02$ & $-0.5$ & $-0.48$ & $-0.480056$ & $-0.4800516$ & $%
-0.4800516$ \\
$0.04$ & $-0.5$ & $-0.46$ & $-0.460448$ & $-0.4603776$ & $-0.460378$ \\
$0.05$ & $-0.5$ & $-0.45$ & $-0.450875$ & $-0.4507031$ & $-0.4507047$ \\
$0.06$ & $-0.5$ & $-0.44$ & $-0.441512$ & $-0.4411556$ & $-0.4411605$ \\
$0.08$ & $-0.5$ & $-0.42$ & $-0.423584$ & $-0.4224576$ & $-0.4224852$ \\
$0.10$ & $-0.5$ & $-0.40$ & $-0.407$ & $-0.40425$ & $-0.4043555$ \\
$0.20$ & $-0.5$ & $-0.30$ & $-0.356$ & $-0.312$ & $-0.318752$ \\
$0.30$ & $-0.5$ & $-0.20$ & $-0.389$ & $-0.16625$ & $-0.2431595$ \\
$\beta $ & $(E_{20}/a^{2})_{0}$ & $(E_{20}/a^{2})_{1}$ & $(E_{20}/a^{2})_{2}$
& $(E_{20}/a^{2})_{4}$ & $(E_{20}/a^{2})_{6}$ \\
\tableline$0.02$ & $-0.2222222$ & $-0.2022222$ & $-0.2024982$ & $-0.2024523$
& $-0.2024526$ \\
$0.04$ & $-0.2222222$ & $-0.1822222$ & $-0.1844302$ & $-0.1836958$ & $%
-0.1837180$ \\
$0.06$ & $-0.2222222$ & $-0.1622222$ & $-0.1696742$ & $-0.1659563$ & $%
-0.1662097$ \\
$0.08$ & $-0.2222222$ & $-0.1422222$ & $-0.1598862$ & $-0.1481358$ & $%
-0.1495594$ \\
$0.10$ & $-0.2222222$ & $-0.1222222$ & $-0.1567222$ & $-0.1280347$ & $%
-0.1334655$%
\end{tabular}
\end{table}

\end{document}